\documentclass{emulateapj}

\usepackage{natbib}
\usepackage{amsmath}

\newcommand{\about}{$\sim\!\!$~}
\newcommand{\kms}{\,km\,s$^{-1}$}

\def\lsim{\hbox{\rlap{\raise 0.425ex\hbox{$<$}}\lower 0.65ex\hbox{$\sim$}}}
\def\gsim{\hbox{\rlap{\raise 0.425ex\hbox{$>$}}\lower 0.65ex\hbox{$\sim$}}}

\def\arcdeg{\mbox{$^\circ$}}

\newcommand{\vsi}{\protect\hbox{$v_{\rm Si~II}$}}
\newcommand{\vsiz}{\protect\hbox{$v_{\rm Si~II}^{0}$}}
\newcommand{\vca}{\protect\hbox{$v_{\rm Ca~H\&K}$}}
\newcommand{\vcaz}{\protect\hbox{$v_{\rm Ca~H\&K}^{0}$}}

\shorttitle{High-$z$ SN~Ia Velocity, Intrinsic Color, and Host Mass}
\shortauthors{Foley}

\begin{document}

 \title{The Relation Between Ejecta Velocity, Intrinsic Color, and Host-Galaxy Mass for High-redshift Type I\lowercase{a} Supernovae}

\def\cfa{1}
\def\clay{2}

\author{
{Ryan~J.~Foley}\altaffilmark{\cfa,\clay}
}

\altaffiltext{\cfa}{
Harvard-Smithsonian Center for Astrophysics,
60 Garden Street, 
Cambridge, MA 02138.
}
\altaffiltext{\clay}{
Clay Fellow. Electronic address rfoley@cfa.harvard.edu .
}

\begin{abstract}
Recently, using a large low-redshift sample of Type Ia supernovae
(SNe~Ia), we discovered a relation between SN~Ia ejecta velocity and
intrinsic color that improves the distance precision of SNe~Ia and
reduces potential systematic biases related to dust reddening.  No
SN~Ia cosmological results have yet made a correction for the
``velocity-color'' relation.  To test the existence of such a relation
and constrain its properties at high redshift, we examine a sample of
75 SNe~Ia discovered and observed by the Sloan Digital Sky Survey-II
(SDSS-II) Supernova Survey and Supernova Legacy Survey (SNLS).  From
each spectrum, we measure ejecta velocities at maximum brightness for
the Ca H\&K and \ion{Si}{2} $\lambda 6355$ features, \vcaz\ and \vsiz,
respectively.  Using SN light-curve parameters, we determine the
intrinsic $B_{\rm max} - V_{\rm max}$ for each SN.  Similar to what
was found at low-redshift, we find that SNe~Ia with higher ejecta
velocity tend to be intrinsically redder than SNe~Ia with lower ejecta
velocity.  The distributions of ejecta velocities for SNe~Ia at low
and high redshift are similar, indicating that current cosmological
results should have little bias related to the velocity-color
relation.  Additionally, we find a slight (2.4-$\sigma$ significant)
trend between SN~Ia ejecta velocity and host-galaxy mass such that
SNe~Ia in high-mass host galaxies tend to have lower ejecta velocities
as probed by \vcaz.  These results emphasize the importance of
spectroscopy for SN~Ia cosmology.
\end{abstract}


\defcitealias{Foley11:vel}{FK11}
\defcitealias{Foley11:vgrad}{FSK11}


\section{Introduction}\label{s:intro}

Type Ia supernovae (SNe~Ia) are reasonable standard candles in the
optical, with a dispersion in their peak magnitudes of \about 0.5~mag
\citep[e.g.,][]{Baade38, Kowal68}.  An empirical relation between the
peak luminosity and light-curve shape \citep{Phillips93} of a SN~Ia
combined with accounting for host-galaxy dust extinction
\citep{Riess96} standardizes SN~Ia luminosities such that after
corrections, the dispersion in peak magnitudes is \about 0.15~mag in
the optical \citep[e.g.,][]{Hicken09:lc}.  Recently, we discovered
that after accounting for the ``width-luminosity'' relation,
low-redshift SNe~Ia with higher-velocity ejecta tend to be
intrinsically redder than those with lower-velocity ejecta
\citep[hereafter FK11]{Foley11:vel}, which can improve SN~Ia distance
precision by a factor of \about 2.  We also presented a method to
correct for this intrinsic color difference when there is a
measurement of either the \ion{Ca}{2} H\&K or \ion{Si}{2} $\lambda
6355$ velocities (\vca\ and \vsi, respectively) near maximum
brightness \citep[hereafter FSK11]{Foley11:vgrad}, which are proxies
for the velocity of all ejecta in the direction of our line of sight.
The combination of the width-luminosity and ``velocity-color''
relations make SNe~Ia standardizable candles and crayons.

The distance precision achieved after correcting for light-curve shape
and observed color is adequate to determine that the expansion of the
Universe is accelerating \citep{Riess98:Lambda, Perlmutter99} and
constrain the equation-of-state parameter of dark energy, $w$
\citep{Wood-Vasey07, Riess07, Hicken09:de, Kessler09:cosmo,
Amanullah10, Conley11, Sullivan11}.  However, neglecting the relation
between ejecta velocity and intrinsic color can potentially bias
cosmological results at a level that will significantly bias
measurements of $w$.

Cosmological parameter measurement with SNe~Ia uses differential
distance measurements, such that assuming an incorrect SN~Ia intrinsic
color will not bias results as long as the intrinsic color
distribution (and particularly the average) does not change with
redshift.  If the average intrinsic color changes with redshift, then
the color offset will result in an incorrect measurement of
host-galaxy extinction, and thus a biased average distance.  Since the
intrinsic color difference between low and high-velocity SNe~Ia is
\about 0.1~mag \citepalias{Foley11:vel}, extremely different low and
high-redshift populations could result in up to a \about 0.15~mag bias
in distance measurements.  Thus far, the velocity-color relation has
not been tested at high redshift.

The velocity-color relation is also linked to several aspects of SN
explosion physics.  First, \vsi\ is strongly correlated with its
velocity gradient, $\dot{v}_{\rm Si~II}$ \citepalias{Foley11:vgrad}.
\citet{Maeda10:asym} found that $\dot{v}_{\rm Si~II}$ and velocity
offset of nebular lines at late times were correlated, indicating an
asymmetric explosion as the origin of the velocity differences.
\citet{Maeda11} found that $\dot{v}_{\rm Si~II}$ was correlated with
intrinsic color, supporting the results of \citetalias{Foley11:vel}.
The polarization of the \ion{Si}{2} $\lambda 6355$ feature, which
probes asymmetry in the outer ejecta, is also correlated with velocity
\citep{Leonard05, Maund10:asym}.  \citet{Ganeshalingam11} found that
SNe~Ia with higher-velocity tend to have shorter $B$-band rise times
than lower-velocity SNe~Ia.

In the Chandrasekhar-mass spherically symmetric explosion model, the
kinetic energy of the ejecta is proportional to the square of the
velocity.  More generally, the kinetic energy per unit mass is
proportional the square of the velocity.  However, once spherical
symmetry is broken, an additional factor relating the measured
velocity to the average velocity is necessary to link kinetic energy
and velocity \citep[e.g.,][]{Kasen07:asym}.  Nonetheless, Velocity
information may be an effective way to answer questions about SN~Ia
progenitor systems and explosion physics.  Measuring these properties
at high redshift has the potential to provide additional information
for understanding SN~Ia explosion physics and the progenitor question.

High-redshift SNe~Ia are relatively faint, and 8-m class telescopes
are typically required to obtain spectra.  There are now hundreds of
high-redshift SN~Ia spectra \citep[e.g.,][]{Coil00, Hook05, Howell05,
Matheson05, Bronder08, Ellis08, Zheng08, Balland09, Foley09:year4,
Foley10:sdss, Konishi11:subaru, Ostman11, Walker11}.  Because of their
faintness and the scarcity of large-aperture telescope time, spectral
time series of high-redshift SNe~Ia are rarely obtained.  Previous
studies have shown that the ensembles of low and high-redshift SNe~Ia
have similar spectral evolution \citep[e.g.,][]{Blondin06, Bronder08,
Foley08:comp}.  Under the assumption that the low and high-redshift
SNe~Ia evolve in a similar fashion, a single near-maximum spectrum is
adequate to examine the velocity-color relation
\citepalias{Foley11:vgrad}.  There have been several high-redshift SN
surveys, but to address the velocity-color relation, which is a
relatively small effect, it is best to use large, homogeneous samples.
It is therefore best to focus on the recent ESSENCE
\citep{Miknaitis07}, Sloan Digital Sky Survey-II Supernovae Survey
\citep[SDSS-II;][]{Frieman08}, and Supernova Legacy Survey
\citep[SNLS;][]{Astier06} programs.

ESSENCE photometry consists of only two bands, and is therefore not
well suited for separating intrinsic color and dust reddening.  The
current SDSS-II sample is relatively small (only 103 SNe~Ia with
published light curves), but has excellent multi-band light curves for
most SNe and because of its intermediate redshift range ($z \lesssim
0.4$), optical spectra typically cover both Ca H\&K and \ion{Si}{2}
$\lambda 6355$.  SNLS has a large published sample of light curves
(242 SNe~Ia with measured light-curve properties) and spectra (330
publicly available spectra).  Additionally, the SNLS SN~Ia redshift
range ($0.4 \lesssim z \lesssim 1$) is higher and complementary to the
redshift range of SDSS-II, which could help see any potential redshift
evolution in the velocity-color relation.

After correcting for light-curve shape and observed color, SN~Ia
absolute magnitudes (or equivalently, Hubble residuals) correlate with
host-galaxy mass, morphology, and star-formation rate
\citep{Hicken09:de, Kelly10, Lampeitl10:host, Sullivan10}.  This
empirical relation does not have an understood physical effect,
although it could be host-galaxy dust properties, progenitor age,
metallicity, or other effects \citep[e.g.,][]{Hoflich98, Lentz00,
Ropke04, Mazzali06, Conley07, Sauer08}.  Since the low and
high-redshift host-galaxy populations have different properties (both
from selection effects and redshift evolution; \citealt{Howell07}), not
accounting for this relation will bias cosmological results
\citep{Kelly10, Lampeitl10:host, Sullivan10}.

In this paper, we perform an analysis of the combined SDSS-II and SNLS
sample of SNe~Ia and compare the sample to the
\citetalias{Foley11:vgrad} low-redshift sample to examine the
velocity-color relation at high redshift.  After making several
practical and quality cuts, we obtain subsamples of the high-redshift
sample with appropriate light-curve properties and well-determined
maximum-brightness ejecta velocity measurements
(Section~\ref{s:samples}).  In Section~\ref{s:vel}, we measure \vca,
\vsi, \vcaz, and \vsiz\ for the high-redshift sample.  In
Section~\ref{s:results}, we find no difference in the average velocity
properties between the low and high-redshift samples; similar to
SNe~Ia in the low-redshift sample, higher-velocity SNe~Ia in the
high-redshift sample are observed to be redder than their
lower-velocity counterparts; SNe~Ia in the high-redshift sample with
higher-velocity ejecta tend to be intrinsically redder, confirming the
velocity-color relation at high redshift; and there is a marginal
trend between ejecta velocity and host-galaxy mass.  We summarize and
conclude in Section~\ref{s:conc}.


\section{Samples}\label{s:samples}

For our study, we use three samples of SNe~Ia.  The first is the
high-redshift SNLS sample; the second is the intermediate-redshift
SDSS-II sample.  We merge the SDSS-II and SNLS samples to generate our
high-redshift sample.  The light curves of both the SDSS-II and SNLS
samples were fit by \citet{Conley11}, who provide a consistent set of
light-curve parameters for all high-redshift SNe.  The last sample is
the low-redshift sample of \citetalias{Foley11:vgrad}.  We discuss the
samples and our quality cuts below.

\subsection{SNLS}

With the release of the 3-year SNLS cosmology results \citep{Guy10,
Conley11, Sullivan11}, the SNLS team also released the one-dimensional
spectra for their sample\footnote{All spectra can be obtained at
https://tspace.library.utoronto.ca/handle/1807/26549 .}.
Most of these spectra have been presented in various publications
\citep{Howell05, Bronder08, Ellis08, Balland09, Walker11}; however,
several spectra have not been published.  Details of the selection,
observations, and spectral reductions can be found in the various SNLS
spectroscopy references.  In total, there are 330 spectra of 316
SNe~Ia.

Although the spectroscopic sample is quite large, it must be
considerably culled for our study.  First, we require light curve
information for our study \citep{Guy10}.  Specifically, times of
maximum brightness, light-curve shape measurements, and some
indication of the SN color are necessary.  Many of the spectra are of
SNe obtained after year 3 of SNLS, for which there are no published
light curves, or have light curves that are not adequate for fitting,
and thus do not have corresponding photometric measurements presented
by \citet{Conley11}.  After matching to the photometry sample, we are
left with 212 spectra of 205 SNe~Ia, or 85\% of the photometry sample.

\citetalias{Foley11:vgrad} was able to correct \vsi\ and \vca\ to
their maximum-light values, \vsiz\ and \vcaz, respectively, if the
spectra were obtained near maximum brightness.  Specifically, the
measurements must be made within $-6 \le t \le 10$~days and $-4 \le t
\le 9$~days in the rest frame for \vsi\ and \vca, respectively.
Additionally, \citetalias{Foley11:vgrad} found that this relation held
for SNe~Ia with $1 \le \Delta m_{15} (B) \le 1.5$~mag.  This
light-curve shape constraint was made to properly match the $\Delta
m_{15} (B)$ dispersion for the low-redshift low and high-velocity
samples \citep{Wang09:2pop}; however, the velocity relations hold for
$0.9 \le \Delta m_{15} (B) \le 1.5$~mag, and we use this range for our
analysis to include more SNe.  Using Equation~5 of \citet{Conley08},
we can convert the $\Delta m_{15} (B)$ ranges with minimum values of 1
and 0.9~mag to stretch ranges of $0.772 \le s \le 1$ and $0.772 \le s
\le 1.067$, respectively.  After making the phase and light-curve
shape cuts, we are left with 106 and 115 spectra of 101 and 110 SNe~Ia
for \vcaz\ and \vsiz, respectively.

Although the SNLS sample of SN~Ia spectra is of relatively good
quality for high-redshift data, not all spectra are adequate for
measuring ejecta velocities.  We perform a visual inspection of all
spectra to determine their fidelity and reject spectra that are noisy,
have large sky residuals, have poor flux calibration, or other obvious
problems.  In addition, because of the redshift range of the sample
and the wavelength range of the spectra, most spectra do not cover
\ion{Si}{2} $\lambda 6355$ and some do not cover Ca H\&K, thus
precluding a measurement of their ejecta velocity from those lines.
Finally, we use the criteria determined by \citetalias{Foley11:vgrad}
for determining if a \vca\ measurement indicates the true ejecta
velocity or that of a high-velocity component and/or blend with
another line (e.g., \ion{Si}{2} $\lambda 3858$; \citealt{Kirshner93};
\citealt{Nugent97}; \citealt{Howell06}).  Specifically, we reject
spectra with a visually apparent red shoulder in the line profile and
spectra that have both $f_{\rm Ca} = f_{\lambda} (v_{\rm Ca~H\&K} +
9000~{\rm km~s}^{-1}) / f_{\lambda} (v_{\rm Ca~H\&K}) < 1.5$ and
$v_{\rm Ca~H\&K} < -14,000$~\kms.  After making the spectral quality
cuts, we are left with 13 and 39 spectra of 13 and 38 SNe~Ia for
\vsiz\ and \vcaz, respectively.

A summary of the cuts and how they affect the sample are presented in
Table~\ref{t:cuts}.

\begin{deluxetable*}{rcccccc}
\tablewidth{0pc}
\tablecaption{High-Redshift Spectroscopy Sample Quality Cuts\label{t:cuts}}
\tablehead{
\colhead{Cut / Sample} & \colhead{Full \vcaz} & \colhead{Full \vsiz} &  \colhead{SDSS \vcaz} & \colhead{SDSS \vsiz} &  \colhead{SNLS \vcaz} & \colhead{SNLS \vsiz}
}

\startdata
 Full SDSS+SNLS Sample        & \multicolumn{2}{c}{511 (447)} & \multicolumn{2}{c}{181 (131)} & \multicolumn{2}{c}{330 (316)} \\
 Has Photometric Measurements & \multicolumn{2}{c}{311 (282)} & \multicolumn{2}{c}{99  (92)}  & \multicolumn{2}{c}{212 (205)} \\
 Light-Curve Shape and Phase  & 159 (145) & 168 (154) & 47 (39) & 53 (44) & 106 (101) & 115 (110) \\
 Spectrum Quality             &  58  (54) &  46  (40) & 19 (16) & 33 (27) &  39  (38) &  13  (13) 

\enddata

\tablecomments{Number of spectra and SNe are presented without and
with parentheses, respectively.}

\end{deluxetable*}

After making all cuts, we are left with 4\% and 12\% of the original
sample for analysis of the \ion{Si}{2} $\lambda 6355$ and Ca H\&K,
respectively.  We use 6\% and 18\% of the sample with photometric
measurements.  If we assume that photometric measurements will be
provided in the future for the remaining sample and that the
percentages measured here are representative of the full sample, then
we can expect \about 6 and \about 18 additional spectra for the two
measurements.  A summary of the cuts and how they affect the sample
are presented in Table~\ref{t:cuts}.

The \ion{Si}{2} $\lambda 6355$ feature is rarely measured for SNe in
the SNLS sample.  This results in the largest reduction for that
feature.  Future surveys which find more SNe~Ia at $z < 0.4$
(approximately where the feature is redshifted beyond the optical
range) should provide a significant number of SNe~Ia where this
feature can be observed with optical spectrographs.  Alternatively,
near-infrared spectroscopy could measure this feature at higher
redshift.  The number of spectra which pass all Ca H\&K cuts is
reduced by both the light-curve shape cut and the quality of the data.
Additional investigations of the velocity evolution as a function of
light-curve shape may help with the former cut.  Longer exposure times
for spectral observations (or larger-aperture telescopes) will help
with the latter cut.

\subsection{SDSS-II}

SDSS-II have released their first-year photometry \citep{Holtzman08},
spectroscopy \citep{Zheng08}, and cosmological results
\citep{Kessler09:cosmo}.  Some of the first-year spectra (along with
additional data) are also presented by \citet{Foley10:sdss} and
\citet{Konishi11:subaru}.  Details of the SN photometry and spectral
reductions can be found in those references.  Details of the SN
detection and selection for spectroscopic follow-up observations can
be found in \citet{Sako08}.  Here, we use the \citet{Zheng08}
compilation, which has 181 spectra of 131 SNe~Ia.

The sample was culled using the same criteria outlined for the SNLS
sample.  Requiring photometric measurements reduced the sample to 99
spectra of 92 SNe~Ia.  Our light-curve shape and phase cut further
reduced the sample to 47 spectra of 39 SNe~Ia and 53 spectra of 44
SNe~Ia for \vcaz\ and \vsiz, respectively.  The quality of the spectra
varies significantly across the sample, but there are many excellent
spectra.  After making the spectroscopy quality cuts above, we are
left with a final SDSS sample of 19 spectra of 16 SNe~Ia and 33
spectra of 27 SNe~Ia for \vcaz\ and \vsiz, respectively.  A summary of
the cuts and how they affect the sample are presented in
Table~\ref{t:cuts}.

\subsection{Combined High-Redshift Sample}

Combining the SDSS-II and SNLS samples, we generate our high-redshift
sample.  This sample consists of 58 spectra with good \vcaz\
measurements for 54 SNe~Ia with appropriate light-curve parameters.
There are also 46 spectra with good \vsiz\ measurements for 40 SNe~Ia
with appropriate light-curve parameters in our high-redshift sample.
There are a total of 75 high-redshift SNe~Ia with at least one of the
two ejecta velocities measured.

\begin{figure*}
\begin{center}
\epsscale{0.8}
\rotatebox{90}{
\plotone{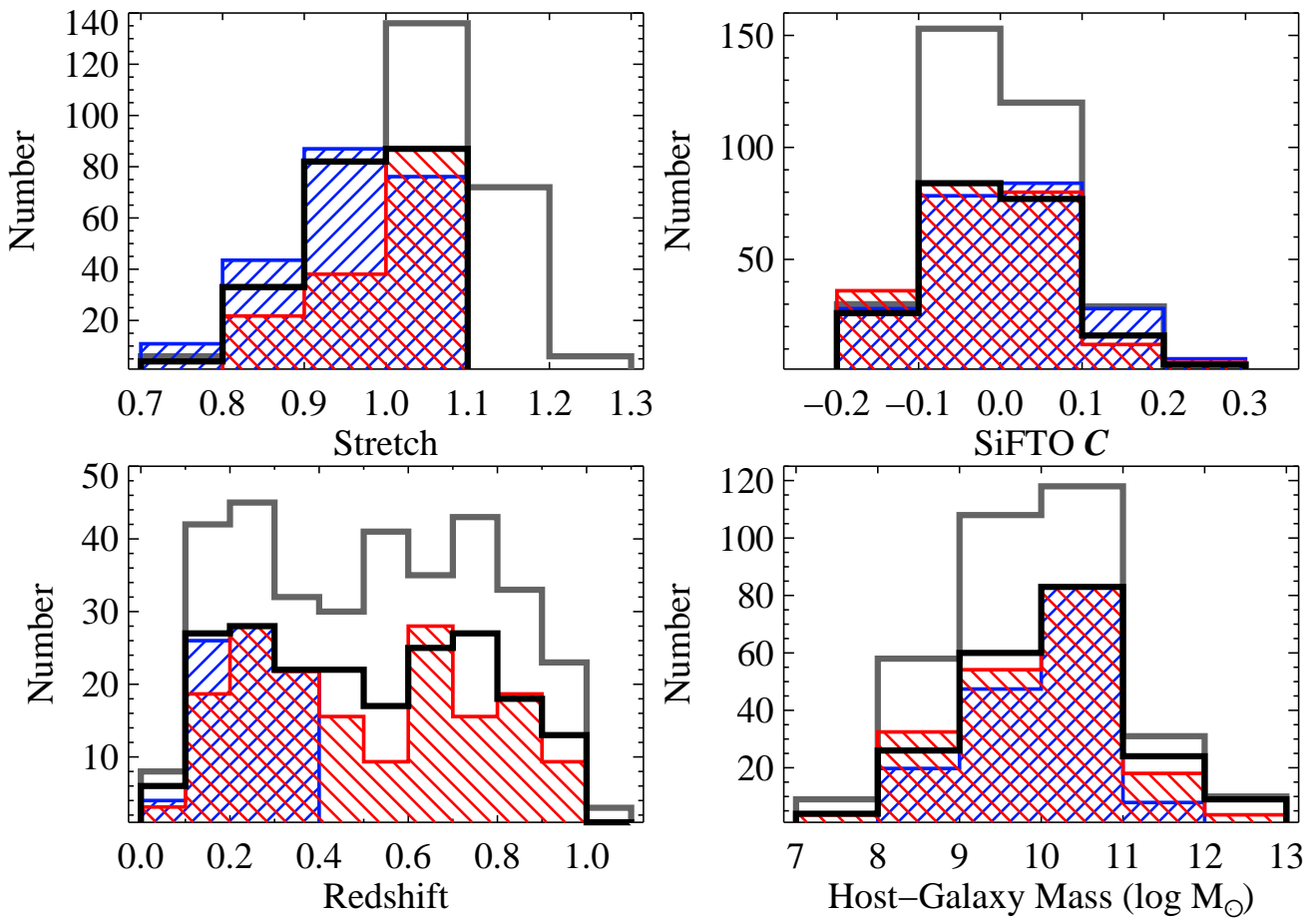}}
\caption{Histograms of stretch (upper left), SiFTO $\mathcal{C}$
(upper right), redshift (lower left), and host-galaxy mass (lower
right) for the high-redshift sample.  The full and light-curve
shape-cut SDSS+SNLS samples are represented by the grey and black
histograms, respectively.  The \vsiz\ and \vcaz\ high-redshift samples
are represented by the blue and red histograms, respectively; their
histograms have been scaled such that their peaks match the peak of
the light-curve shape-cut high-redshift histogram.}\label{f:hist}
\end{center}
\end{figure*}

We present distributions of the stretch, SiFTO $\mathcal{C}$
\citep{Conley08}, redshift, and host-galaxy mass for the full
high-redshift, light-curve shape-cut high-redshift, \vsiz\
high-redshift, and \vcaz\ high-redshift samples in
Figure~\ref{f:hist}.  In that figure, the \vsiz\ high-redshift and
\vcaz\ high-redshift samples were scaled such that the peaks of their
histograms match the peaks of the light-curve shape-cut high-redshift
sample histograms.  Comparing the subsamples using Kolmogorov-Smirnov
(K-S) tests shows that the subsamples with velocity information are
mostly consistent with being drawn from the same parent population as
the light-curve shape-cut high-redshift sample (all have $p$-values
$>$0.1).  The single exception is the \vsiz\ sample in regard to
redshift, which has a K-S $p$ value of $7 \times 10^{-11}$.  The
results of the K-S tests are similar when comparing the velocity
subsamples to the full SNLS+SDSS sample except the stretch
distributions are significantly different ($p < 1 \times 10^{-5}$).
Therefore, the velocity subsamples appear to be reasonable
representations of the light-curve shape-cut high-redshift sample.

A table of all high-redshift SNe/spectra which pass our cuts, their
light-curve measurements, and their ejecta velocity measurements can
be found in Table~\ref{t:data}.

\subsection{FSK11}

\citetalias{Foley11:vgrad} presented photometric and spectroscopic
properties for a large sample of low-redshift SNe~Ia.  The
spectroscopic sample was largely taken from the CfA Supernova Program
(\citealt{Matheson08}; S.\ Blondin et~al., in preparation).  The
photometry was taken from the CfA3 \citep{Hicken09:lc} and LOSS
\citep{Ganeshalingam11} samples.  In total, the sample relies on 1630
optical spectra of 255 SNe~Ia.  The final sample consists of 42 and 65
SNe~Ia which passed all quality cuts and have measurements for \vcaz
and \vsiz, respectively.


\section{Ejecta Velocity}\label{s:vel}

Using the method of \citet{Blondin06}, we measure \vsi\ and \vca\ for
the high-redshift sample.  Briefly, the spectra are deredshifted and
smoothed using an inverse-variance-weighted Gaussian filter.  The
smoothed spectra are resampled on a fine wavelength scale with
wavelength bins of 0.1~\AA.  The minimum of each feature
(corresponding to the wavelength of maximum absorption) is recorded,
and a velocity is measured using the relativistic Doppler formula.
The uncertainty in the velocity measurements include the uncertainty
of determining the wavelength of maximum absorption, differences for
assuming different smoothing scales, and redshift uncertainties.  We
exclude spectra that are of low quality and \vca\ measurements
according to the criteria of \citetalias{Foley11:vgrad} and outlined
above.

The velocity inferred from the minimum of a spectral feature does not
necessarily correspond to the ``photospheric'' velocity of the
ejecta.  If the line is saturated, the velocity measured from that
line will be higher than that of the photospheric velocity.  However,
if the density profiles of the element responsible for the the
spectral line is smooth with radius, the velocity inferred from the
line is physically meaningful and a good proxy for the overall ejecta
kinematics.  The Ca H\&K and \ion{Si}{2} $\lambda 6355$ features both
have smooth velocity evolution with time for most SNe over the phase
range we use in this study \citepalias{Foley11:vgrad}, and therefore,
\vsi\ and \vca\ are good proxies for the overall ejecta velocity.

High-redshift SN~Ia spectra tend to have much more galaxy
contamination than those from SNe~Ia at low redshift
\citep{Foley08:comp}.  High-redshift SN~Ia spectra also tend to have
a lower signal-to-noise ratio (S/N).  We examined the possibility of
lower S/N and higher galaxy-contamination biasing velocity
measurements.  Performing a Monte Carlo simulation which added various
amounts of noise and galaxy contamination (from several different
galaxy templates) to several SN~Ia spectra found that there is a
slight bias of \about 200~\kms\ to lower velocities with extreme
galaxy contamination and relatively low S/N.  Visual inspection of the
spectra in the high-redshift sample that pass all of our cuts
indicates that none have significant galaxy contamination, and thus
any bias should be significantly less than our overall measurement
uncertainty.

\begin{figure}
\begin{center}
\epsscale{1.02}
\rotatebox{90}{
\plotone{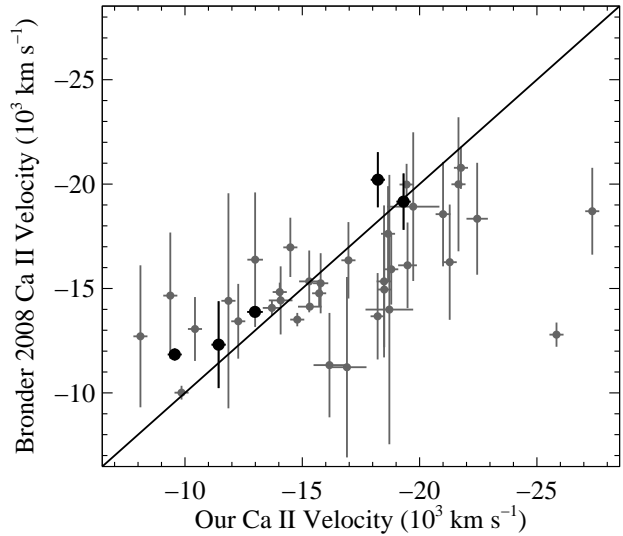}}
\caption{\vca\ measurements from \citet{Bronder08} compared to the
measurements presented here.  The grey symbols are for all spectra for
which a velocity was measured for both studies, including those that
do not pass our quality cuts.  The black symbols are for the 5 spectra
from \citet{Bronder08} that pass all of our cuts.}\label{f:bronder}
\end{center}
\end{figure}

The SNLS has performed several spectroscopic studies, but only one,
\citet{Bronder08}, provided ejecta velocity measurements.
\citet{Bronder08} only measure \vca, but did so by fitting a Gaussian
to a 60-\AA\ smoothed spectrum.  This method is different from our
method, and can provide significantly different velocity measurements
if there are multiple components to the Ca H\&K feature.
Figure~\ref{f:bronder} compares the velocity measured by
\citet{Bronder08} to our measured velocity.  The grey symbols are for
all spectra with measurements in both samples, while the black symbols
are for the spectra which pass all of our cuts.  There is general
agreement between the two studies; however for the spectra that do not
pass all of our cuts, the \citet{Bronder08} measurements span a
smaller velocity range for the full sample.  This is likely the result
of the Gaussian fitting, which will tend to find an intermediate
velocity if there are two velocity components.  Our quality cuts
typically reject such spectra (even if it provides a more accurate
measurement of the velocity).  All spectra which pass our quality cuts
have consistent velocity measurements from both studies.

The SDSS-II survey also has several spectroscopic studies, but none
present tabulated line velocities for a direct comparison.

As a SN photosphere recedes in velocity space with time, the ejecta
velocity as measured from spectral features also decreases.  Since
spectra are obtained at various phases, a correction for this temporal
evolution must be made to properly compare different SNe.
\citetalias{Foley11:vgrad} provides a prescription for making this
correction.  Using the time of maximum brightness measured from the
light curves, along with the measurements of \vsi\ and \vca, the
high-redshift ejecta velocities are corrected to maximum brightness
--- \vsiz\ and \vcaz\ --- using Equations~5 and 7 of
\citetalias{Foley11:vgrad}.  Uncertainties for these measurements
include both the velocity uncertainties and the scatter in the
relations.  These measurements are provided in Table~\ref{t:data}.

There are a handful of SNe with multiple spectra which pass our phase
and quality cuts.  Each of these spectra provide an independent
measurement of the maximum-brightness ejecta velocity.  In these
cases, we choose the higher-quality spectrum to provide the adopted
measurement.  In cases where the difference in quality is not obvious,
we choose the spectrum closer to maximum brightness, as was done by
\citetalias{Foley11:vgrad}.

A table of all high-redshift SNe/spectra, light-curve measurements,
and ejecta velocity measurements can be found in Table~\ref{t:data}.

All measurements for the \citetalias{Foley11:vgrad} sample are
presented there.


\section{Results}\label{s:results}

\subsection{Velocity Distributions}

The \citetalias{Foley11:vel} relation between ejecta velocity and
light-curve shape has significant implications for SN~Ia cosmology.
First, measuring ejecta velocity should improve the precision and
accuracy of SN~Ia distance measurements.  However, since cosmological
measurements are made from differential SN~Ia distances, ignoring this
effect biases cosmological results if either the low and high-redshift
samples have different velocity distributions (and thus different
intrinsic color distributions), or if the relation evolves with
redshift.

If the intrinsic color distributions are different such that the
average intrinsic color is different for low and high redshift, then
the differential extinction correction will be incorrect.  To test
this possibility, we compare the velocity distributions of the
\citetalias{Foley11:vgrad} and high-redshift samples.  The
\citetalias{Foley11:vgrad} sample may not completely overlap with the
\citet{Conley11} low-redshift cosmology sample, but it is
representative of the observed low-redshift population.  Similarly,
the culled high-redshift sample presented here should be
representative of the full high-redshift sample (with the same
light-curve parameters).  To mitigate any possible correlation between
highly reddened SNe~Ia (which are not found at high redshift) and
velocity, we remove all SNe~Ia with $B_{\rm max} - V_{\rm max} >
0.4$~mag from the \citetalias{Foley11:vgrad} sample.

\begin{figure}
\begin{center}
\epsscale{1.15}
\rotatebox{90}{
\plotone{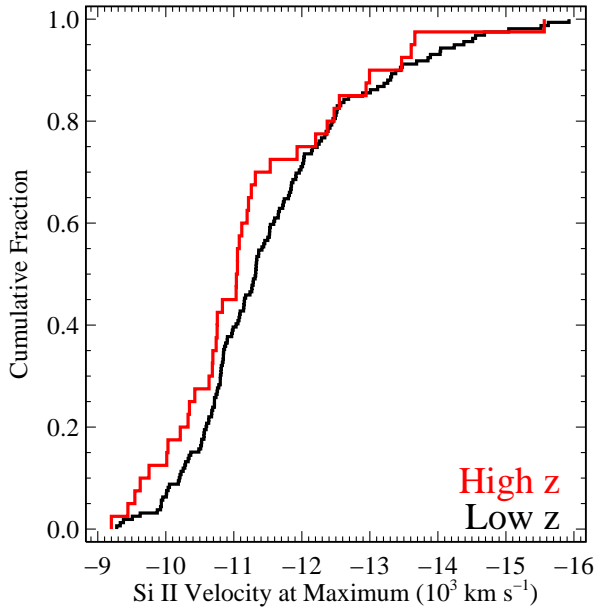}}
\caption{\vsiz\ CDF for the \citetalias{Foley11:vgrad} (black) and
high-redshift (red) samples.}\label{f:si_cdf}
\end{center}
\end{figure}

\begin{figure}
\begin{center}
\epsscale{1.15}
\rotatebox{90}{
\plotone{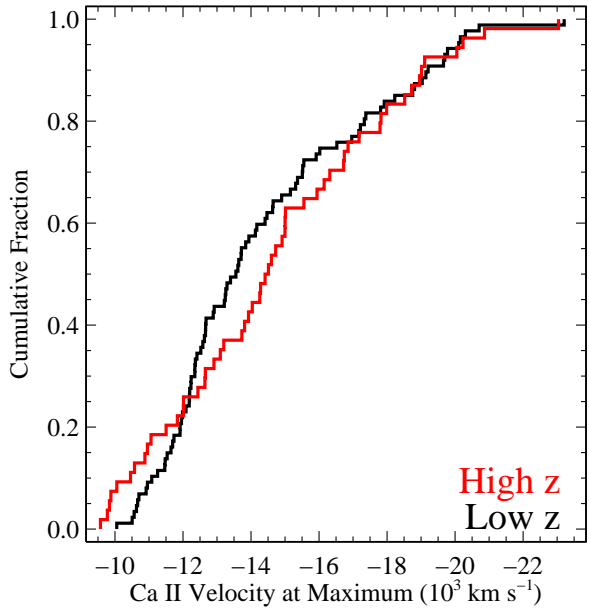}}
\caption{\vcaz\ CDF for the \citetalias{Foley11:vgrad} (black) and
high-redshift (red) samples.}\label{f:ca_cdf}
\end{center}
\end{figure}

Figures~\ref{f:si_cdf} and \ref{f:ca_cdf} present the \vsiz\ and
\vcaz\ cumulative distribution functions (CDFs), respectively, for the
\citetalias{Foley11:vgrad} and high-redshift samples.  The two samples
appear relatively similar for both quantities.  K-S tests result in
$p$-values of 0.10 and 0.20 for \vsiz\ and \vcaz, respectively.  From
these tests, there is no indication that the samples are pulled from
different parent populations in regard to ejecta velocity.

There is no significant velocity offset between the low and
high-redshift samples.  As will be shown in Section~\ref{ss:velcol},
ejecta velocity and intrinsic color are correlated for our
high-redshift sample.  Assuming that the relations between ejecta
velocity and intrinsic color are the same at low and high redshift and
since the low and high-redshift samples have similar ejecta velocity
distributions, it is unlikely that the low and high-redshift samples
have significantly different intrinsic colors.  Therefore, it is
unlikely that cosmological measurements are highly biased by not
accounting for the velocity-color relation.  However, the
high-redshift samples are still small, and even a small offset could
affect cosmological measurements.  Future surveys and studies should
consider the possibility of a difference in low and high-redshift
SN~Ia ejecta velocities.

\subsection{Correlation Between Ejecta Velocity and Color}\label{ss:velcol}

\citetalias{Foley11:vel} showed that SNe~Ia with high ejecta velocity
(as determined by \ion{Si}{2} $\lambda 6355$) have redder intrinsic
$B_{\rm max} - V_{\rm max}$ colors than SNe~Ia with low ejecta
velocity.  \citetalias{Foley11:vgrad} showed this relation held when
using Ca H\&K to determine ejecta velocity and further showed that a
linear function adequately describes the relation between intrinsic
$B_{\rm max} - V_{\rm max}$ ($(B_{\rm max} - V_{\rm max})_{0}$) and
\vsiz.  Here, we test the relation between ejecta velocity and
intrinsic color for the high-redshift sample.  If this relation
evolves with redshift, cosmological measurements can be significantly
affected.

Unfortunately, there are no published measurements of $B_{\rm max} -
V_{\rm max}$ for the high-redshift sample.  However, there are
measurements of $\mathcal{C}$, the color parameter from SiFTO
\citep{Conley08}.  SiFTO does not attempt to distinguish intrinsic
color and dust extinction, and parameterizes differences in apparent
color as a single parameter, $\mathcal{C}$.  Examining this parameter
is similar to examining an observed $B_{\rm max} - V_{\rm max}$.  The
high-redshift sample consists of SNe~Ia with little to no dust
reddening.  Therefore, using $\mathcal{C}$ as a proxy for intrinsic
color is a reasonable first approximation.

\begin{figure}
\begin{center}
\epsscale{1.0}
\rotatebox{90}{
\plotone{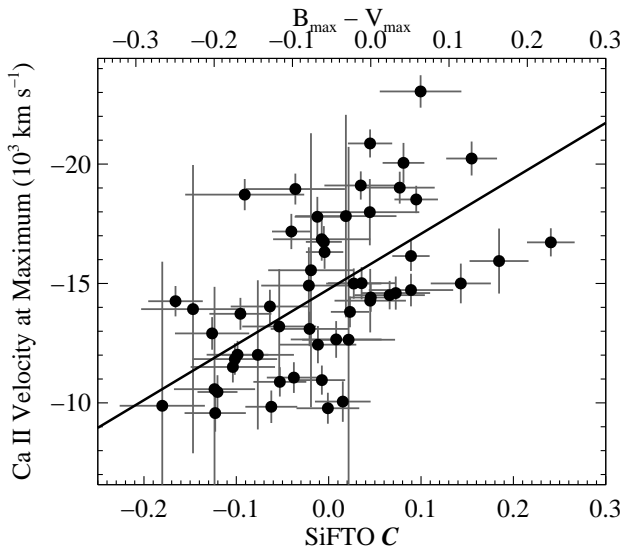}}
\caption{Maximum-light \ion{Ca}{2} H\&K velocity (\vca) as a function
of SiFTO $\mathcal{C}$.  The equivalent values for $B_{\rm max} -
V_{\rm max}$ as determined by Equation~\ref{e:bvc} are presented on
the top axis.  The solid line represents the best-fit linear model for
the data.}\label{f:velc}
\end{center}
\end{figure}

Figure~\ref{f:velc} presents \vcaz\ as a function of $\mathcal{C}$ for
the high-redshift sample.  Performing a Bayesian Monte-Carlo linear
regression on the data \citep{Kelly07} with $2 \times 10^{6}$
realizations, we find that 99.995\% of the realizations have a
negative slope (all but 11), which corresponds to a 4.0-$\sigma$
result.  The linear correlation coefficient is $-0.61$.  There is a
strong relation between ejecta velocity and observed color for the
high-redshift sample, with the higher-velocity SNe~Ia having redder
colors.  This is similar to what was found for the low-redshift sample
\citep{Pignata08:02dj, Wang09:2pop}.

Performing the same analysis with \vsiz, we find a worse linear
correlation coefficient of $-0.28$ and the linear trend is less than a
2-$\sigma$ result.  This difference may be the result of the small
sample or possibly the relatively small number of SNe~Ia with
high-velocity \ion{Si}{2} $\lambda 6355$ (11/40 SNe~Ia with a \vsiz\
measurement have velocities above $-11,800$~\kms), preventing a strong
linear trend.  Future larger samples will determine if the
low-significance trend seen with \vsiz\ and $\mathcal{C}$ is real at
high redshift.

\begin{figure}
\begin{center}
\epsscale{1.0}
\rotatebox{90}{
\plotone{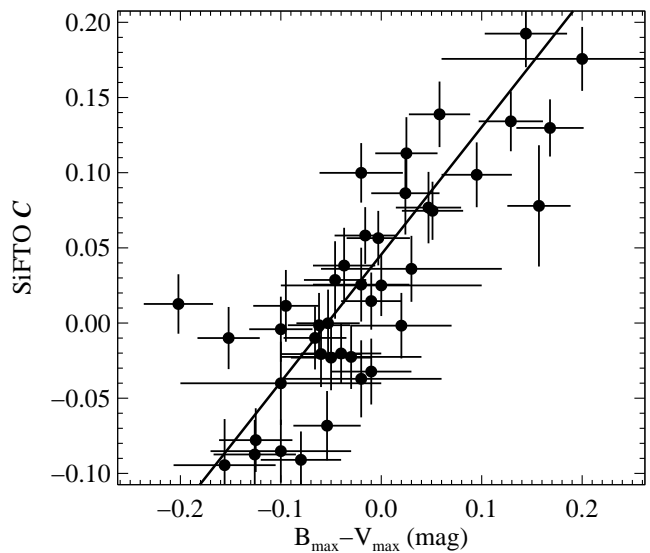}}
\caption{$B_{\rm max} - V_{\rm max}$ and SiFTO $\mathcal{C}$
measurements for the 41 low-redshift SNe~Ia with measurements from
\citet{Conley11} and \citetalias{Foley11:vgrad} and with $0.9 \le
\Delta m_{15} (B) \le 1.5$~mag.  The line represents the best linear
fit to the data.}\label{f:bvc}
\end{center}
\end{figure}

As mentioned above, $\mathcal{C}$ is a reasonable proxy for the
observed SN color.  Using the intersection of the low-redshift sample
used by \citet{Conley11} and the \citetalias{Foley11:vgrad} sample, we
are able to determine a relation between $\mathcal{C}$ and $B_{\rm
max} - V_{\rm max}$.  The measurements for the 41 SNe~Ia in both
samples with our adopted $\Delta m_{15} (B)$ range are presented in
Figure~\ref{f:bvc}.  The two quantities are highly correlated and a
linear fit to the data provides the following relation:
\begin{equation}\label{e:bvc}
  B_{\rm max} - V_{\rm max} = (1.18 \pm 0.11) \mathcal{C} - (0.054 \pm 0.008).
\end{equation}
If one uses all 65 SNe~Ia in the matched sample, the relation is
statistically equivalent.  Since the two quantities are highly
correlated, \vcaz\ is also highly correlated with $B_{\rm max} -
V_{\rm max}$ for the high-redshift sample.

Using the measurement of the observed rest-frame $B$-band maximum
brightness, $m_{B}$, along with the measured stretch and a distance
modulus, $\mu$, we can determine the light-curve shape corrected peak
absolute $B$ magnitude, $M_{B}^{\prime}$.  The choice of cosmological
parameters for determining the distance modulus will slightly change
the value of $M_{B}^{\prime}$.  We choose to use the best-fit
\citet{Sullivan11} Flat $w$CDM values of ($\alpha$, $\Omega_{m}$, $w)
= (1.451$, 0.269, $-1.061$).  We also use $H_{0} =
70.5$~km~s$^{-1}$~Mpc$^{-1}$ to match the analyses of
\citetalias{Foley11:vel} and \citetalias{Foley11:vgrad}, but the
choice of $H_{0}$ only affects the overall normalization.  Choosing a
Flat $\Lambda$CDM cosmology with $\Omega_{m} = 0.3$ makes little
difference on the results of our analysis.  We further adjust the
absolute magnitude to account for the correlation between Hubble
residuals and host-galaxy mass \citep{Kelly10, Lampeitl10:host,
Sullivan10}.  SNe~Ia hosted in high mass galaxies ($M >
10^{10}$~M$_{\sun}$) tend to be brighter by 0.05 -- 0.1~mag after
correction for both stretch and $\mathcal{C}$.  We use the magnitude
difference of \citealt{Sullivan11} of $\Delta \mathcal{M} = 0.076$~mag
to further correct $M_{B}$.  Specifically, $M_{B} = M_{B}^{\prime}$
and $M_{B} = M_{B}^{\prime} + \Delta \mathcal{M}$ for SNe~Ia hosted in
galaxies with $M \le 10^{10}$ and $M > 10^{10}$~M$_{\sun}$,
respectively.  We have performed our analysis both with and without
this correction, and it makes only a slight difference in the
significance of our results.

\begin{figure}
\begin{center}
\epsscale{1.15}
\rotatebox{90}{
\plotone{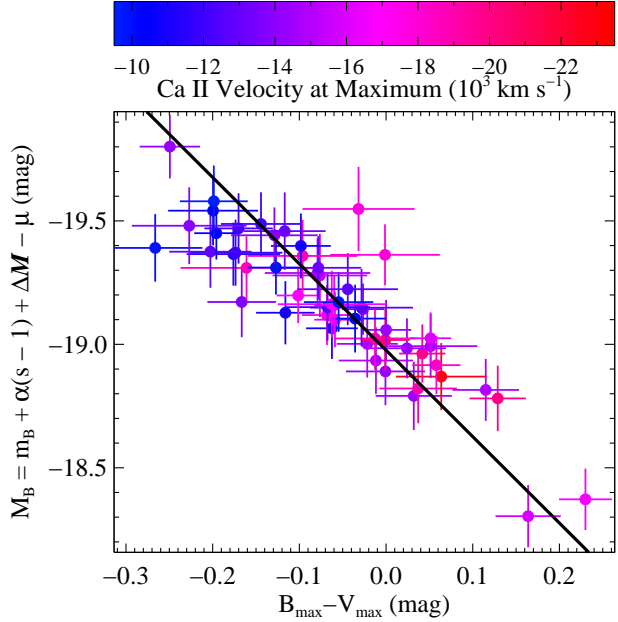}}
\caption{The light-curve shape and host-mass corrected peak absolute
$B$ brightness as a function of $B_{\rm max} - V_{\rm max}$.  The
color of the symbol corresponds to the SN's \vcaz, with the color bar
at the top of the figure displaying the correspondence.  The solid
black line represents the host-galaxy dust reddening of a zero-color
SN~Ia.  The deviations from this line in color is the intrinsic color
of a given SN.}\label{f:cacolmag}
\end{center}
\end{figure}

\begin{figure}
\begin{center}
\epsscale{1.15}
\rotatebox{90}{
\plotone{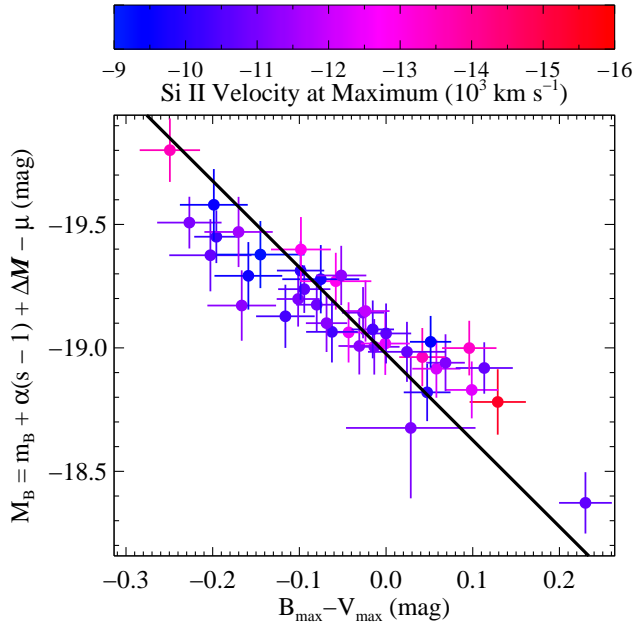}}
\caption{Same as Figure~\ref{f:cacolmag}, except for the high-redshift
sample with \vsiz\ measurements and with the symbols color-coded by
\vsiz.}\label{f:sicolmag}
\end{center}
\end{figure}

Figures~\ref{f:cacolmag} and \ref{f:sicolmag} present $M_{B}$ as a
function of $B_{\rm max} - V_{\rm max}$ for the high-redshift sample.
These figures are similar to Figure~19 from
\citetalias{Foley11:vgrad}.  The symbols are color-coded by \vcaz\ and
\vsiz\ for Figures~\ref{f:cacolmag} and \ref{f:sicolmag},
respectively.  There is a general trend that redder SNe are also
fainter.  Since we have already corrected for light-curve shape, this
is presumably the result of dust reddening.  We overplot the reddening
law found by \citet{Foley11:vel} for the low-velocity
sample\footnote{The reddening law was slightly adjusted to account for
the high-redshift data using $M_{B}$, while the low-redshift samples
used $M_{V}$.  This required accounting for $R_{B} = R_{V} + 1$ and a
slight offset in the magnitude of zero color.}, which matches the
general trend mentioned above.  This reddening law is extremely
similar to the color-correction trend $\beta \mathcal{C}$ used by
SiFTO assuming the best-fit value for $\beta$ from \citet{Sullivan11}.
We have performed our analysis using both relations, and our results
do not change if we choose a particular relation.

\begin{figure}
\begin{center}
\epsscale{1.15}
\rotatebox{90}{
\plotone{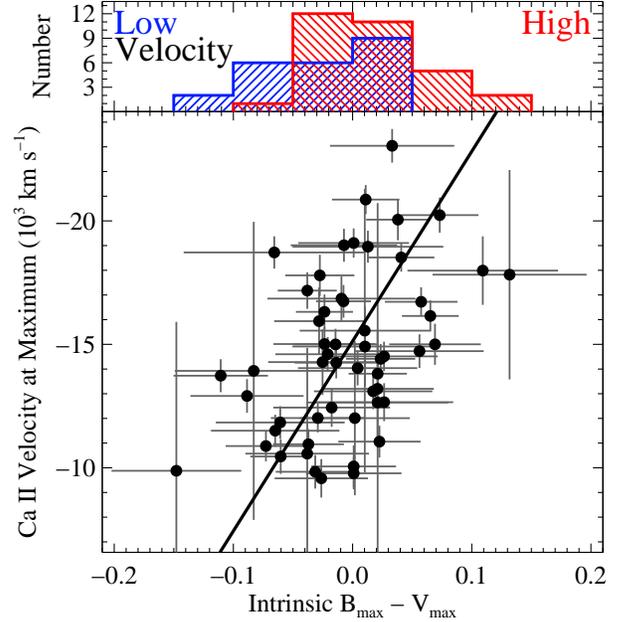}}
\caption{Maximum-light \ion{Ca}{2} H\&K velocity (\vcaz) as
a function of intrinsic $B_{\rm max} - V_{\rm max}$ pseudo-color for
the high-redshift sample of SNe~Ia.  The solid line is the best-fit
linear model for the data.  The histograms (top panel) show the
intrinsic $B_{\rm max} - V_{\rm max}$ pseudo-color distribution for
subsamples split by $v_{\rm Ca~H\&K}^{0} = -14,000$~\kms, with the
blue and red histograms representing the low and high-velocity
subsamples, respectively.}\label{f:cavelcol}
\end{center}
\end{figure}

\begin{figure}
\begin{center}
\epsscale{1.15}
\rotatebox{90}{
\plotone{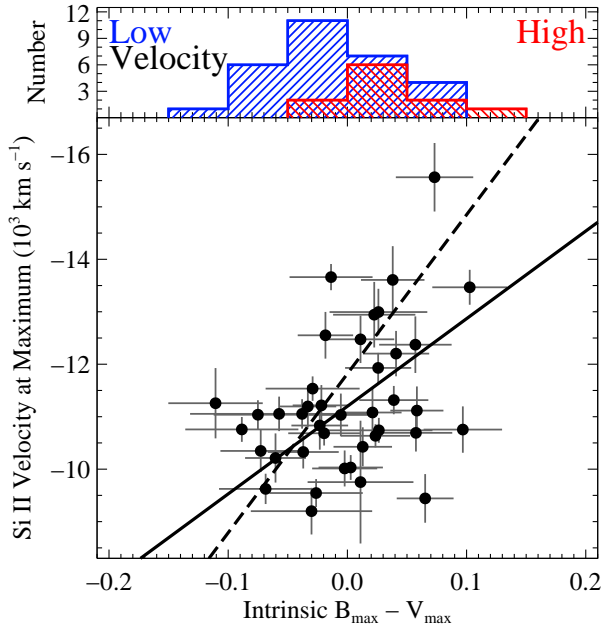}}
\caption{Maximum-light \ion{Si}{2} $\lambda 6355$ velocity (\vsiz) as
a function of intrinsic $B_{\rm max} - V_{\rm max}$ pseudo-color for
the high-redshift sample of SNe~Ia.  The solid line is the best-fit
linear model for the data.  The dashed line is the linear relation
found for the low-redshift sample \citepalias{Foley11:vgrad}.  The
histograms (top panel) show the intrinsic $B_{\rm max} - V_{\rm max}$
pseudo-color distribution for subsamples split by $v_{\rm Si~II}^{0} =
-11,800$~\kms, with the blue and red histograms representing the low
and high-velocity subsamples, respectively.}\label{f:sivelcol}
\end{center}
\end{figure}

As was done by \citetalias{Foley11:vgrad}, we use the color offset
from the dust reddening line to determine the intrinsic $B_{\rm max} -
V_{\rm max}$ pseudo-color.  Figures~\ref{f:cavelcol} and
\ref{f:sivelcol} display \vcaz\ and \vsiz\ as a function of intrinsic
$B_{\rm max} - V_{\rm max}$ for the high-redshift sample.  As was seen
in the low-redshift sample \citepalias{Foley11:vel, Foley11:vgrad},
SNe~Ia with higher-velocity ejecta tend to be intrinsically redder.
The color and velocity have a linear correlation coefficients of
$-0.71$ and $-0.53$ for \vcaz\ and \vsiz, respectively, indicating
good correlations.  The correlation between \vcaz\ and intrinsic color
is much better for the high-redshift sample than what was found for
the low-redshift sample \citepalias[$\rho = -0.24$;][]{Foley11:vgrad},
but similar for \vsiz.  Performing a Bayesian Monte-Carlo linear
regression on the \vcaz\ data \citep{Kelly07}, we find that 99.8\% of
the realizations have a negative slope, corresponding to a
3.1-$\sigma$ result.  The best-fit relation is
\begin{align}
  (B_{\rm max} - V_{\rm max})_{0} &= (-0.20 \pm 0.06) - (0.013 \pm 0.004) \notag  \\
    &\times (v_{\rm Ca~H\&K}^{0} / 1000 {\rm ~km~s}^{-1}) {\rm ~mag}.\label{e:bvvel}
\end{align}
It is somewhat surprising that the trend is statistically significant
since a linear trend was not significant for the low-redshift sample
\citepalias{Foley11:vgrad}.

Performing the same analysis for \vsiz, we find that 98.8\% of the
Monte Carlo realizations have a negative slope, corresponding to only
a 2.5-$\sigma$ result.  Although this does not qualify as a
statistically significant result, the best-fit line is similar to that
of the low-redshift sample.  The relatively small number of
high-velocity events (11/40 SNe~Ia with a
\vsiz\ measurement have velocities above $-11,800$~\kms) likely
contributes to the low significance.  Additional high-redshift data
should be able to determine if the linear relation is real.

Along with the linear analysis, we examine the \vcaz\ and \vsiz\
samples after splitting them into low and high-velocity (using
$-14,000$ and $-11,800$~\kms\ for \vcaz\ and \vsiz, respectively).  As
shown by the histograms in Figures~\ref{f:cavelcol} and
\ref{f:sivelcol}, the low and high-velocity subsamples are offset in
intrinsic color.  The low/high-velocity subsamples have mean $(B_{\rm
max} - V_{\rm max})_{0}$ pseudo-colors of $-0.028 \pm 0.008$ / $0.013
\pm 0.006$ and $-0.005 \pm 0.006$ / $0.030 \pm 0.009$~mag for the
\vcaz\ and \vsiz\ samples, respectively.  For both measurements of the
ejecta velocity, there is an \about0.04~mag offset between the low and
high-velocity average intrinsic colors that is significant at the 3.3
/ 4.5-$\sigma$ level.  However, any linear relation between $B_{\rm
max} - V_{\rm max}$ and \vcaz\ should underestimate the statistical
difference by simply separating the sample into two velocity groups.

Finally, we perform K-S tests on the low and high-velocity subsamples.
When comparing the velocity subsamples, the K-S $p$-values for the
\vcaz\ and \vsiz\ samples are $5.0 \times 10^{-3}$ and $8.9 \times
10^{-3}$, respectively.  The K-S tests therefore indicate that the low
and high-velocity subsamples are drawn from different parent
populations in regard to intrinsic color.

An offset of 0.04~mag in $B_{\rm max} - V_{\rm max}$, the difference
in the median values for the low and high-velocity subsamples,
corresponds to a difference in $A_{V}$ of 0.08 to 0.12~mag for $2 \le
R_{V} \le 3.1$.  The high-redshift sample has an intrinsic color
dispersion, $\sigma_{(B_{\rm max} - V_{\rm max})_{0}}$, of 0.051 and
0.050~mag for \vcaz\ and \vsiz, respectively, but splitting the sample
by velocity results in $\sigma_{(B_{\rm max} - V_{\rm max})_{0}} =
0.048$ (0.050) and 0.044 (0.036) mag for the low and high-velocity
\vcaz\ (\vsiz) subsamples, respectively.  This is similar to what was
found for the \citetalias{Foley11:vgrad} sample; however, unlike the
\citetalias{Foley11:vgrad} sample, the low-velocity high-redshift
sample does not have a significantly smaller color scatter than the
high-velocity sample.

\subsection{Ejecta Velocity and Host-Galaxy Mass}

It has long been known that SNe~Ia in different hosts have different
peak luminosities \citep[e.g.,][]{Branch96, Hamuy96:lum, Howell01}.
The observed correlation between SN~Ia host-galaxy mass, morphology,
and star-formation rate and Hubble residuals \citep{Hicken09:de,
Kelly10, Lampeitl10:host, Sullivan10} improves SN~Ia distance
measurements and points to a possible connection between SN~Ia
progenitor properties and SN luminosity that are not already removed
by accounting for light-curve shape.  The secondary effect is such
that SNe~Ia hosted in galaxies with $M \ge 10^{10}$~M$_{\sun}$ are
0.05 -- 0.1~mag brighter than those with the same light-curve
properties hosted in less massive galaxies after making initial
corrections.  Since the intrinsic color difference between low and
high-velocity SNe~Ia can cause an incorrect distance measurement, and
thus absolute magnitude measurement, of the same order of magnitude,
it is prudent to determine if SN~Ia ejecta velocity is correlated with
host-galaxy mass.

The correlation between ejecta velocity and host-galaxy mass has not
been investigated at any redshift.  To increase our sample, we combine
the high-redshift sample with the low-redshift
\citetalias{Foley11:vgrad} sample matched to the \citet{Conley11}
sample to provide consistent host-galaxy mass measurements.  This
increases our sample by 14 and 23 SNe~Ia for our \vcaz\ and \vsiz\
samples.  Several SNe~Ia have reported host-galaxy masses equal to the
maximum mass in the \citet{Conley11} sample.  These SNe either had no
host-galaxy photometry available to the SNLS team or saturated the
galaxy model at the high-mass end.  Since most were low-redshift SNe,
they were assigned to the high-mass bin and the probability that they
were low mass was taken as a systematic uncertainty (A.\ Conley,
private communication).  The lack of an accurate mass estimate for
these galaxies reduces their usefulness for our study, and we
therefore exclude these SNe from our analysis.  In the total combined
sample, there are 67 and 63 SNe~Ia with \vcaz\ and
\vsiz\ measurements, respectively.

\begin{figure*}
\begin{center}
\epsscale{0.75}
\rotatebox{90}{
\plotone{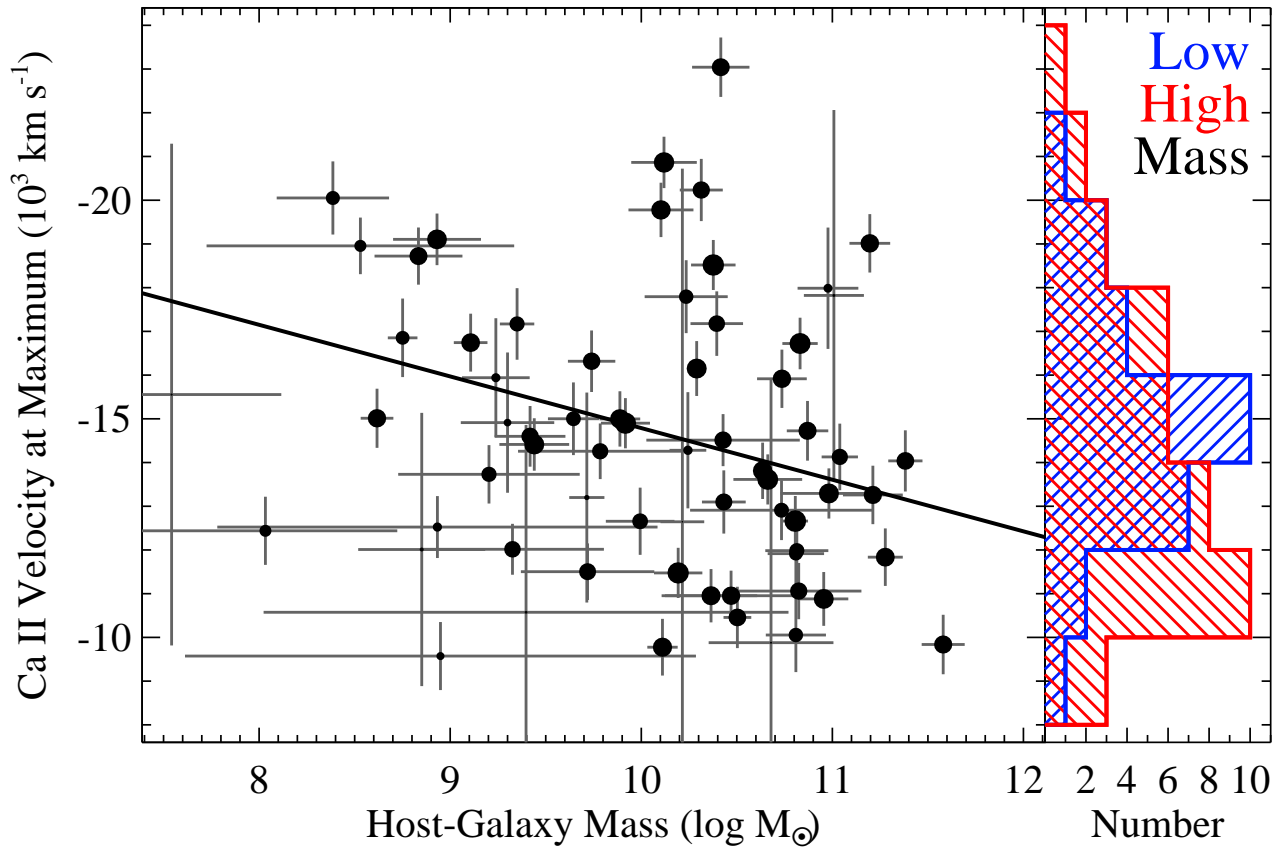}}
\caption{Left panel: \vcaz\ for the \vcaz\ high-redshift sample of
SNe~Ia as a function of host-galaxy mass.  The size of the symbols is
inversely proportional to the size of the uncertainties.  The best-fit
linear model to the data is shown as a solid line.  Right panel:
\vcaz\ histograms for the high-redshift SNe~Ia hosted in galaxies with
masses above/below $10^{10}$~M$_{\sun}$ (red/blue
histograms).}\label{f:camass}
\end{center}
\end{figure*}

Comparing \vcaz\ to host-galaxy mass, there is a reasonable linear
correlation ($\rho = 0.31$), and a Bayesian Monte-Carlo linear
regression results in 98.4\% of the realizations having a positive
slope (a 2.4-$\sigma$ result).  Figure~\ref{f:camass} displays \vcaz\
as a function of host-galaxy mass.  There is a slight visible trend
such that more massive galaxies tend to host SNe~Ia with lower \vcaz.
It also appears that SNe~Ia in high-mass galaxies have a larger spread
in \vcaz.

One worry is that the relations between host-galaxy mass, stretch, and
intrinsic color may cause the observed relation.  Similar to what was
found by \citetalias{Foley11:vgrad}, there is no correlation between
light-curve shape and either ejecta velocity or intrinsic color for
the high-redshift sample ($\rho = -0.066$ for stretch and
\vcaz; $\rho = -0.070$ for stretch and $(B_{\rm max} - V_{\rm
max})_{0}$).  The correlation between stretch and \vcaz\ is slightly
larger for the combined low and high redshift sample ($\rho =
-0.111$), but still not significant.  The correlation between
host-galaxy mass and stretch does not appear to be causing the
correlation between host-galaxy mass and intrinsic color for our
sample.

There is no significant linear relation between host-galaxy mass and
\vcaz\ for our sample, but the significance level warrants a further
investigation with larger data sets.  Splitting the sample at
$10^{10}$~M$_{\sun}$, the low and high-mass samples have average
\vcaz\ of $-15,050 \pm 140$ and $-14,370 \pm 110$~\kms, respectively.
Using Equation~\ref{e:bvvel}, the velocity offset corresponds to an
intrinsic color offset of 0.009~mag.  If this is incorrectly
attributed to dust, it would correspond to an error in a distance
estimate of 0.018 to 0.027~mag for $2 \le R_{V} \le 3.1$.

There is no significant relation between \vsiz\ and host-galaxy mass
for our sample.  Since \vcaz\ is strongly correlated with intrinsic
color and potentially correlated with host-galaxy mass, one might
expect that intrinsic color and host-galaxy mass are correlated.  We
performed this analysis on the full \citet{Conley11} sample and find
no significant correlation.


\section{Discussion \& Conclusions}\label{s:conc}

Using the high-redshift sample, we identified 40 and 54 high-redshift
SNe~Ia with appropriate light-curve shapes, spectral phases, and
spectral data quality to measure the maximum-light Ca H\&K and
\ion{Si}{2} $\lambda 6355$ ejecta velocities (\vcaz\ and \vsiz),
respectively.  We compare the distributions of ejecta velocity for the
high-redshift sample to the distributions of the low-redshift
\citetalias{Foley11:vgrad} sample, finding no statistically significant
difference in the distributions for the two samples.  We measure the
intrinsic $B_{\rm max} - V_{\rm max}$ pseudo-color for the
high-redshift sample.  The ejecta velocity is highly correlated with
$(B_{\rm max} - V_{\rm max})_{0}$, such that SNe~Ia with higher ejecta
velocity tend to be redder.  This is similar to what has been shown
for the low-redshift sample \citepalias{Foley11:vel, Foley11:vgrad}.
We compare ejecta velocity to host-galaxy mass, finding a slight trend
between the quantities such that SNe~Ia hosted in lower-mass galaxies
tend to have higher ejecta velocities than those hosted in higher-mass
galaxies.  The significance of this relation is still low, and future
studies should examine this relation further.

A larger data set is required to determine if the average SN~Ia ejecta
velocity changes with redshift.  This is particularly true for \vsiz,
which is hampered by its relatively red wavelength.  Near-infrared
spectroscopy could increase the sample of SNe~Ia with \vsiz\
measurements at high redshift.

The confirmation of the velocity-color relation at high redshift is an
important step towards better cosmological measurements.  Although the
velocity-color relation affects some light-curve properties
\citep{Ganeshalingam11, Cartier11}, it is still unclear if one can
infer the ejecta velocity or intrinsic color of SNe~Ia from their
light curves alone.  Therefore, the most precise SN~Ia distance
measurements at all redshifts continue to require spectroscopy.

The asymmetric explosion model suggests that the distribution of
measured ejecta velocities can be explained by asymmetric explosions
and viewing explosions from different viewing angles
\citep{Maeda10:asym, Foley11:vel}.  If this model is correct, our
measured distribution of ejecta velocities for low and high-redshift
SNe~Ia indicate that approximately the same amount of asymmetry is
present at low and high redshifts, further indicating that SN~Ia
explosions do not evolve significantly from $z = 0$ to $z \approx
0.5$.  Regardless of symmetry, this result also indicates that the
distribution of kinetic energy per unit mass is similar across this
redshift range.

The tentative relation between ejecta velocity and host-galaxy mass
may, if confirmed, provide insight into the relation between
host-galaxy mass and Hubble residuals.  Ejecta velocity is more
directly connected to the SN properties than the host-galaxy mass, and
might help the search for the physical driver of this relation.

Additional high-redshift data from SDSS-II and SNLS, which has already
been obtained, will significantly increase this sample.  Current and
future high-redshift SN surveys such as Pan-STARRS, the Dark Energy
Survey, and eventually the Large Synoptic Survey Telescope, will
provide large samples of high-redshift SNe~Ia.  However, spectroscopy
from other telescopes is required to further examine the relations
between ejecta velocity, intrinsic color, and host-galaxy properties
at high redshift.

\begin{acknowledgments} 

\bigskip
R.J.F.\ is supported by a Clay Fellowship.

This work was started and completed while R.J.F.\ was visiting the
Carnegie Observatories in Pasadena.  He thanks J.\ Simon, J.\
Kollmeier, and E.\ Teague for facilitating this extremely stimulating
visit.  He thanks N.\ Sanders for providing comments and A.\ Conley
for helping determine the fidelity of some of the data.

\end{acknowledgments}

\bibliographystyle{fapj}
\bibliography{../astro_refs}

\clearpage
\LongTables
\begin{deluxetable*}{lrrrrrrrrrr}
\tablewidth{0pc}
\tablecaption{High-Redshift Spectroscopy Sample\label{t:data}}
\tablehead{
\colhead{} & \colhead{} & \colhead{Phase} & \colhead{} & \colhead{} & \colhead{$m_{B}$} & \colhead{Host Mass} & \colhead{\vca} & \colhead{\vsi} & \colhead{\vcaz} & \colhead{\vsiz} \\
\colhead{SN} & \colhead{Redshift} & \colhead{(days)} & \colhead{Stretch} & \colhead{SiFTO $\mathcal{C}$} & \colhead{(mag)} & \colhead{($\log {\rm M}_{\sun}$)} & \colhead{($10^{3}$~\kms)} & \colhead{($10^{3}$~\kms)} & \colhead{($10^{3}$~\kms)} & \colhead{($10^{3}$~\kms)}
}

\startdata
SDSS2165 & 0.2880 (0.0050) & $ 6.19$ (0.33) & 1.06 (0.05) & $-0.096$ (0.034) & 21.604 (0.043) &  9.203 (0.475) & $-13.28$ (0.35) & $-10.78$ (0.63) & $-13.73$ (0.67) & $-11.26$ (0.67) \\
SDSS2372 & 0.1810 (0.0005) & $ 2.13$ (0.20) & 1.03 (0.03) & $ 0.045$ (0.024) & 20.580 (0.035) & 10.119 (0.171) & $-19.93$ (0.14) & $-12.23$ (0.39) & $-20.86$ (0.59) & $-12.47$ (0.45) \\
SDSS2561 & 0.1180 (0.0002) & $ 0.01$ (0.15) & 0.99 (0.02) & $ 0.086$ (0.023) & 19.813 (0.033) & 10.574 (0.086) & \nodata         & $-10.03$ (0.09) & \nodata         & $-10.03$ (0.24) \\
SDSS2561\tablenotemark{a} & 0.1180 (0.0002) & $ 0.01$ (0.15) & 0.99 (0.02) & $ 0.086$ (0.023) & 19.813 (0.033) & 10.574 (0.086) & \nodata         & $ -9.10$ (0.25) & \nodata         & $ -9.10$ (0.33) \\
SDSS2789 & 0.2900 (0.0005) & $ 3.93$ (0.33) & 0.90 (0.05) & $-0.077$ (0.043) & 21.576 (0.056) & 10.949 (0.191) & \nodata         & $ -9.16$ (0.39) & \nodata         & $ -9.20$ (0.44) \\
SDSS2916\tablenotemark{a} & 0.1240 (0.0005) & $-2.32$ (0.14) & 0.88 (0.03) & $ 0.066$ (0.038) & 19.937 (0.062) & 10.428 (0.401) & $-15.08$ (0.20) & $-11.02$ (0.19) & $-14.78$ (0.61) & $-10.87$ (0.29) \\
SDSS2916 & 0.1240 (0.0005) & $ 1.24$ (0.14) & 0.88 (0.03) & $ 0.066$ (0.038) & 19.937 (0.062) & 10.428 (0.401) & $-14.37$ (0.17) & $-10.67$ (0.09) & $-14.51$ (0.60) & $-10.74$ (0.24) \\
SDSS2992 & 0.1270 (0.0001) & $ 8.87$ (0.20) & 0.89 (0.02) & $ 0.127$ (0.027) & 20.034 (0.046) &  9.881 (0.301) & \nodata         & $-12.15$ (0.25) & \nodata         & $-13.47$ (0.33) \\
SDSS3080 & 0.1740 (0.0005) & $-4.51$ (0.12) & 1.00 (0.02) & $-0.038$ (0.023) & 20.236 (0.031) & 10.710 (0.030) & \nodata         & $-10.18$ (0.26) & \nodata         & $-10.01$ (0.34) \\
SDSS3087 & 0.1650 (0.0005) & $-1.26$ (0.12) & 1.06 (0.02) & $ 0.025$ (0.024) & 20.266 (0.032) &  9.373 (0.089) & \nodata         & $-12.05$ (0.18) & \nodata         & $-11.93$ (0.29) \\
SDSS3451 & 0.2500 (0.0005) & $-3.28$ (0.22) & 1.06 (0.03) & $-0.038$ (0.029) & 20.958 (0.037) & 10.824 (0.328) & $-10.85$ (0.30) & $-13.38$ (0.59) & $-11.06$ (0.64) & $-12.94$ (0.63) \\
SDSS3592\tablenotemark{a} & 0.0870 (0.0002) & $-3.51$ (0.08) & 0.97 (0.01) & $-0.040$ (0.021) & 18.751 (0.026) & 10.395 (0.138) & $-18.34$ (0.26) & $-11.35$ (0.17) & $-17.42$ (0.63) & $-11.10$ (0.28) \\
SDSS3592 & 0.0870 (0.0002) & $ 0.17$ (0.08) & 0.97 (0.01) & $-0.040$ (0.021) & 18.751 (0.026) & 10.395 (0.138) & $-17.14$ (0.47) & $-11.04$ (0.17) & $-17.18$ (0.74) & $-11.05$ (0.28) \\
SDSS5533 & 0.2200 (0.0005) & $ 2.29$ (0.23) & 0.98 (0.03) & $ 0.046$ (0.025) & 21.173 (0.034) &  9.440 (0.183) & $-14.17$ (0.19) & $-10.50$ (0.02) & $-14.41$ (0.60) & $-10.64$ (0.22) \\
SDSS5549 & 0.1210 (0.0050) & $ 0.29$ (0.12) & 1.02 (0.01) & $ 0.033$ (0.021) & 19.654 (0.026) &  8.749 (0.132) & \nodata         & $-10.41$ (0.30) & \nodata         & $-10.43$ (0.37) \\
SDSS5844 & 0.3110 (0.0005) & $ 5.03$ (0.36) & 1.01 (0.05) & $-0.099$ (0.033) & 21.571 (0.051) &  9.326 (0.479) & $-12.09$ (0.12) & $-11.10$ (0.08) & $-12.02$ (0.58) & $-11.54$ (0.23) \\
SDSS5957 & 0.2800 (0.0005) & $ 2.81$ (0.26) & 0.98 (0.05) & $-0.089$ (0.033) & 21.453 (0.043) & 10.232 (0.147) & \nodata         & $ -9.55$ (0.19) & \nodata         & $ -9.62$ (0.29) \\
SDSS6057 & 0.0670 (0.0001) & $ 3.36$ (0.12) & 0.94 (0.02) & $ 0.129$ (0.026) & 18.641 (0.044) &  9.944 (0.075) & \nodata         & $-11.99$ (0.49) & \nodata         & $-12.37$ (0.54) \\
SDSS6192 & 0.2720 (0.0005) & $ 5.69$ (0.38) & 0.83 (0.06) & $-0.018$ (0.038) & 21.698 (0.046) &  9.611 (0.567) & \nodata         & $ -9.59$ (1.15) & \nodata         & $ -9.75$ (1.17) \\
SDSS6304 & 0.1900 (0.0005) & $ 6.04$ (0.20) & 0.93 (0.04) & $ 0.095$ (0.024) & 20.952 (0.028) & 10.377 (0.116) & $-16.59$ (0.04) & $-11.55$ (0.37) & $-18.52$ (0.57) & $-12.20$ (0.43) \\
SDSS6315\tablenotemark{a} & 0.2670 (0.0005) & $-0.30$ (0.25) & 0.97 (0.04) & $-0.166$ (0.030) & 20.919 (0.044) &  9.785 (0.430) & $-14.71$ (0.47) & $-12.13$ (0.33) & $-14.67$ (0.74) & $-12.10$ (0.40) \\
SDSS6315 & 0.2670 (0.0005) & $ 2.66$ (0.25) & 0.97 (0.04) & $-0.166$ (0.030) & 20.919 (0.044) &  9.785 (0.430) & $-13.99$ (0.28) & $-13.25$ (0.12) & $-14.26$ (0.64) & $-13.66$ (0.25) \\
SDSS6699\tablenotemark{a} & 0.3110 (0.0005) & $-2.38$ (0.34) & 0.87 (0.07) & $-0.126$ (0.040) & 21.796 (0.049) & 10.734 (0.479) & \nodata         & $-10.18$ (1.10) & \nodata         & $-10.08$ (1.12) \\
SDSS6699 & 0.3110 (0.0005) & $ 3.72$ (0.34) & 0.87 (0.07) & $-0.126$ (0.040) & 21.796 (0.049) & 10.734 (0.479) & $-12.79$ (0.38) & $-10.53$ (0.09) & $-12.91$ (0.68) & $-10.76$ (0.24) \\
SDSS6780 & 0.2020 (0.0050) & $ 9.10$ (0.29) & 0.79 (0.04) & $-0.004$ (0.035) & 20.947 (0.056) & 11.170 (0.221) & \nodata         & $-11.78$ (0.38) & \nodata         & $-12.99$ (0.44) \\
SDSS6933 & 0.2130 (0.0050) & $ 0.44$ (0.17) & 0.99 (0.03) & $ 0.002$ (0.025) & 20.832 (0.030) &  8.975 (0.866) & \nodata         & $-11.28$ (0.16) & \nodata         & $-11.32$ (0.27) \\
SDSS6936 & 0.1810 (0.0005) & $-0.34$ (0.21) & 1.00 (0.03) & $-0.007$ (0.025) & 20.575 (0.031) & 10.365 (0.242) & $-10.93$ (0.22) & $-10.34$ (0.22) & $-10.95$ (0.61) & $-10.33$ (0.31) \\
SDSS7147\tablenotemark{a} & 0.1100 (0.0005) & $-1.32$ (0.14) & 0.80 (0.02) & $-0.034$ (0.026) & 19.516 (0.030) & 10.310 (0.023) & \nodata         & $ -7.79$ (0.04) & \nodata         & $ -7.83$ (0.22) \\
SDSS7147 & 0.1100 (0.0005) & $ 0.40$ (0.14) & 0.80 (0.02) & $-0.034$ (0.026) & 19.516 (0.030) & 10.310 (0.023) & \nodata         & $-10.66$ (0.08) & \nodata         & $-10.68$ (0.23) \\
SDSS7475 & 0.3220 (0.0050) & $ 3.07$ (0.33) & 1.02 (0.06) & $-0.123$ (0.033) & 21.535 (0.042) &  8.949 (1.337) & $-10.00$ (0.53) & $ -9.48$ (0.15) & $ -9.57$ (0.78) & $ -9.54$ (0.27) \\
SDSS7512 & 0.2190 (0.0050) & $-1.58$ (0.23) & 1.05 (0.04) & $ 0.027$ (0.030) & 21.104 (0.044) &  9.888 (0.106) & $-15.22$ (0.28) & \nodata         & $-15.00$ (0.63) & \nodata         \\
SDSS7847 & 0.2120 (0.0005) & $-1.42$ (0.24) & 1.02 (0.04) & $ 0.155$ (0.028) & 21.225 (0.033) & 10.314 (0.112) & $-20.81$ (0.43) & $-15.87$ (0.61) & $-20.23$ (0.71) & $-15.56$ (0.65) \\
SDSS8046 & 0.2590 (0.0005) & $-3.32$ (0.30) & 1.05 (0.05) & $ 0.077$ (0.038) & 21.633 (0.044) & 11.196 (0.106) & $-20.16$ (0.35) & $-10.34$ (0.11) & $-19.01$ (0.67) & $-10.19$ (0.24) \\
SDSS10434 & 0.1040 (0.0005) & $ 2.52$ (0.22) & 1.01 (0.03) & $-0.053$ (0.029) & 19.185 (0.042) & 10.955 (0.128) & $-11.06$ (0.23) & $-10.23$ (0.34) & $-10.88$ (0.62) & $-10.35$ (0.40) \\
SDSS11864 & 0.3030 (0.0005) & $ 3.62$ (0.79) & 1.01 (0.17) & $ 0.070$ (0.063) & 22.299 (0.076) &  9.270 (0.301) & \nodata         & $-10.80$ (0.24) & \nodata         & $-11.05$ (0.33) \\
  03D1ax & 0.4960 (0.0010) & $-2.17$ (0.09) & 0.93 (0.01) & $-0.062$ (0.028) & 22.992 (0.035) & 11.580 (0.112) & $ -9.56$ (0.22) & $ -6.44$ (0.00) & $ -9.84$ (0.68) & $ -6.59$ (0.37) \\
  03D1dt & 0.6120 (0.0020) & $ 7.26$ (0.69) & 1.05 (0.07) & $-0.054$ (0.040) & 23.300 (0.047) &  9.715 (0.092) & $-12.86$ (2.25) & \nodata         & $-13.20$ (2.40) & \nodata         \\
  03D1ew & 0.8680 (0.0010) & $ 2.15$ (0.40) & 1.04 (0.04) & $-0.036$ (0.054) & 24.359 (0.072) &  8.530 (0.805) & $-18.22$ (0.06) & \nodata         & $-18.95$ (0.65) & \nodata         \\
  03D3bl & 0.3553 (0.0005) & $ 3.58$ (0.32) & 1.00 (0.03) & $ 0.241$ (0.026) & 22.951 (0.032) & 10.831 (0.093) & $-15.91$ (0.00) & $-10.48$ (0.24) & $-16.72$ (0.59) & $-10.69$ (0.36) \\
  03D4at & 0.6340 (0.0010) & $ 5.84$ (0.38) & 1.02 (0.04) & $-0.008$ (0.053) & 23.733 (0.047) &  8.751 (0.078) & $-15.49$ (0.63) & \nodata         & $-16.85$ (0.90) & \nodata         \\
  03D4cx & 0.9490 (0.0140) & $ 2.76$ (0.55) & 0.95 (0.07) & $ 0.019$ (0.055) & 24.464 (0.082) & 11.008 (0.157) & $-17.04$ (0.22) & \nodata         & $-17.82$ (4.24) & \nodata         \\
  04D1dc & 0.2110 (0.0010) & $-1.71$ (0.21) & 0.86 (0.02) & $ 0.023$ (0.021) & 21.084 (0.026) & 10.635 (0.041) & $-13.94$ (0.02) & $-11.20$ (0.27) & $-13.81$ (0.64) & $-11.08$ (0.46) \\
  04D1hx & 0.5600 (0.0020) & $ 5.29$ (0.20) & 1.04 (0.02) & $ 0.143$ (0.033) & 23.715 (0.039) &  9.645 (0.133) & $-14.27$ (0.08) & \nodata         & $-15.00$ (0.83) & \nodata         \\
  04D1ow\tablenotemark{a} & 0.9150 (0.0140) & $ 3.56$ (0.18) & 1.00 (0.03) & $-0.124$ (0.044) & 24.366 (0.070) &  9.397 (1.373) & $ -9.38$ (0.29) & \nodata         & $ -8.72$ (4.25) & \nodata         \\
  04D1ow & 0.9150 (0.0140) & $ 6.06$ (0.18) & 1.00 (0.03) & $-0.124$ (0.044) & 24.366 (0.070) &  9.397 (1.373) & $-11.11$ (0.65) & \nodata         & $-10.57$ (4.29) & \nodata         \\
  04D1pd & 0.9500 (0.0200) & $ 1.92$ (0.35) & 1.04 (0.05) & $ 0.022$ (0.050) & 24.734 (0.078) & 10.215 (0.115) & $-12.61$ (5.38) & \nodata         & $-12.64$ (8.08) & \nodata         \\
  04D1pp & 0.7350 (0.0010) & $ 1.68$ (0.20) & 0.87 (0.03) & $-0.064$ (0.042) & 23.998 (0.049) & 11.382 (0.090) & $-13.89$ (0.27) & \nodata         & $-14.04$ (0.70) & \nodata         \\
  04D1sk & 0.6634 (0.0005) & $-0.58$ (0.31) & 1.03 (0.04) & $ 0.100$ (0.044) & 24.058 (0.046) & 10.416 (0.150) & $-23.36$ (0.34) & \nodata         & $-23.04$ (0.68) & \nodata         \\
  04D2an & 0.6200 (0.0190) & $-3.52$ (0.46) & 0.99 (0.08) & $-0.019$ (0.046) & 23.597 (0.046) &  7.542 (0.574) & $-16.14$ (0.40) & \nodata         & $-15.55$ (5.74) & \nodata         \\
  04D2cf & 0.3690 (0.0020) & $ 7.74$ (1.36) & 0.88 (0.05) & $ 0.015$ (0.030) & 22.491 (0.065) & 10.809 (0.157) & $-10.95$ (0.19) & \nodata         & $-10.05$ (0.85) & \nodata         \\
  04D2fp & 0.4150 (0.0020) & $ 2.13$ (0.15) & 1.03 (0.02) & $-0.012$ (0.025) & 22.559 (0.034) & 10.235 (0.218) & $-17.19$ (0.11) & \nodata         & $-17.79$ (0.83) & \nodata         \\
  04D2fs & 0.3570 (0.0020) & $ 1.99$ (0.14) & 1.01 (0.02) & $ 0.081$ (0.023) & 22.437 (0.031) &  8.386 (0.294) & $-19.26$ (0.13) & $-13.30$ (0.08) & $-20.05$ (0.84) & $-13.61$ (0.64) \\
  04D2gp & 0.7320 (0.0010) & $ 2.84$ (0.36) & 0.85 (0.04) & $-0.091$ (0.064) & 24.219 (0.063) &  8.834 (0.230) & $-17.79$ (0.11) & \nodata         & $-18.72$ (0.65) & \nodata         \\
  04D2mc & 0.3480 (0.0010) & $ 6.31$ (0.11) & 0.84 (0.02) & $ 0.142$ (0.028) & 22.580 (0.040) & 10.960 (0.118) & \nodata         & $-10.37$ (0.24) & \nodata         & $-10.75$ (0.44) \\
  04D3cp & 0.8300 (0.0200) & $ 5.74$ (0.17) & 1.05 (0.03) & $-0.180$ (0.046) & 24.111 (0.060) & 10.679 (0.326) & $-10.59$ (0.36) & \nodata         & $ -9.88$ (6.03) & \nodata         \\
  04D3ez & 0.2630 (0.0005) & $ 1.29$ (0.12) & 0.89 (0.02) & $ 0.089$ (0.020) & 21.697 (0.026) & 10.290 (0.012) & $-15.89$ (0.22) & $ -9.42$ (0.38) & $-16.15$ (0.63) & $ -9.44$ (0.46) \\
  04D3mk & 0.8130 (0.0010) & $-1.69$ (0.18) & 0.95 (0.03) & $-0.104$ (0.046) & 24.294 (0.058) &  9.719 (0.349) & $-11.43$ (0.07) & \nodata         & $-11.50$ (0.65) & \nodata         \\
  04D3nh & 0.3402 (0.0002) & $ 4.50$ (0.10) & 1.06 (0.02) & $ 0.009$ (0.020) & 22.142 (0.029) &  9.356 (0.021) & \nodata         & $-12.02$ (0.38) & \nodata         & $-12.55$ (0.44) \\
  04D4ju & 0.4720 (0.0010) & $-2.63$ (0.22) & 1.05 (0.03) & $ 0.184$ (0.032) & 23.771 (0.035) &  9.239 (0.177) & $-16.43$ (1.20) & \nodata         & $-15.94$ (1.36) & \nodata         \\
  05D1cb & 0.6320 (0.0010) & $ 5.09$ (0.23) & 0.97 (0.03) & $-0.001$ (0.034) & 23.715 (0.043) & 10.111 (0.079) & $-10.43$ (0.07) & \nodata         & $ -9.78$ (0.65) & \nodata         \\
  05D1ke & 0.6900 (0.0100) & $ 2.60$ (0.21) & 1.04 (0.03) & $-0.077$ (0.039) & 23.611 (0.044) &  8.851 (0.333) & $-12.05$ (0.67) & \nodata         & $-12.01$ (3.12) & \nodata         \\
  05D2ab & 0.3230 (0.0010) & $-1.26$ (0.14) & 0.99 (0.02) & $-0.013$ (0.020) & 22.001 (0.030) & 10.098 (0.080) & \nodata         & $-11.29$ (0.15) & \nodata         & $-11.20$ (0.40) \\
  05D2ah & 0.1840 (0.0010) & $-2.08$ (0.10) & 0.99 (0.02) & $ 0.019$ (0.020) & 20.765 (0.025) &  9.210 (0.014) & \nodata         & $-11.37$ (0.11) & \nodata         & $-11.22$ (0.39) \\
  05D2ci & 0.6300 (0.0010) & $ 7.42$ (0.17) & 0.90 (0.03) & $ 0.045$ (0.054) & 23.612 (0.058) & 10.977 (0.159) & $-15.82$ (1.23) & \nodata         & $-17.99$ (1.39) & \nodata         \\
  05D2dt & 0.5740 (0.0040) & $-1.61$ (0.22) & 1.02 (0.04) & $ 0.045$ (0.039) & 23.656 (0.047) & 10.244 (0.096) & $-14.44$ (0.05) & \nodata         & $-14.28$ (1.33) & \nodata         \\
  05D2he & 0.6075 (0.0005) & $ 4.15$ (0.33) & 1.04 (0.04) & $ 0.073$ (0.037) & 23.953 (0.044) &  9.418 (0.185) & $-14.11$ (0.37) & \nodata         & $-14.60$ (0.69) & \nodata         \\
  05D2le & 0.7002 (0.0005) & $ 6.89$ (0.22) & 1.05 (0.03) & $-0.012$ (0.042) & 23.961 (0.050) &  8.033 (0.690) & $-12.39$ (0.51) & \nodata         & $-12.44$ (0.78) & \nodata         \\
  05D2nn & 0.8700 (0.0200) & $-0.29$ (0.34) & 0.88 (0.06) & $-0.147$ (0.056) & 24.395 (0.086) & 12.073 (0.487) & $-13.95$ (0.40) & \nodata         & $-13.93$ (6.04) & \nodata         \\
  05D3lb & 0.6473 (0.0004) & $-0.61$ (0.18) & 1.04 (0.02) & $ 0.035$ (0.039) & 23.896 (0.043) &  8.931 (0.230) & $-19.32$ (0.09) & \nodata         & $-19.10$ (0.59) & \nodata         \\
  05D3mn & 0.7600 (0.0010) & $ 2.80$ (0.22) & 0.99 (0.03) & $-0.021$ (0.042) & 24.043 (0.048) & 10.432 (0.115) & $-12.98$ (0.33) & \nodata         & $-13.10$ (0.72) & \nodata         \\
  05D3mq & 0.2460 (0.0010) & $ 8.30$ (0.15) & 0.91 (0.03) & $ 0.034$ (0.025) & 21.521 (0.032) & 10.723 (0.076) & \nodata         & $-10.45$ (0.17) & \nodata         & $-11.03$ (0.41) \\
  05D3ne & 0.1692 (0.0002) & $-4.01$ (0.09) & 0.81 (0.03) & $-0.147$ (0.031) & 20.251 (0.041) & 11.133 (0.162) & \nodata         & $-11.32$ (0.16) & \nodata         & $-11.04$ (0.28) \\
  05D4cw & 0.3750 (0.0010) & $ 7.25$ (0.14) & 0.91 (0.01) & $-0.120$ (0.022) & 22.145 (0.031) & 10.503 (0.072) & $-11.14$ (0.28) & $ -9.89$ (0.28) & $-10.46$ (0.70) & $-10.21$ (0.47) \\
  05D4dw & 0.8550 (0.0010) & $ 5.11$ (0.31) & 1.05 (0.04) & $ 0.008$ (0.049) & 24.438 (0.071) &  9.994 (0.179) & $-12.56$ (0.42) & \nodata         & $-12.66$ (0.77) & \nodata         \\
  05D4fg & 0.8390 (0.0010) & $ 0.39$ (0.22) & 1.02 (0.03) & $-0.102$ (0.046) & 24.195 (0.058) & 11.277 (0.091) & $-11.84$ (0.13) & \nodata         & $-11.84$ (0.66) & \nodata         \\
  05D4fo & 0.3730 (0.0010) & $-5.65$ (0.07) & 0.92 (0.01) & $-0.022$ (0.020) & 22.463 (0.031) &  8.381 (0.366) & \nodata         & $-11.19$ (0.27) & \nodata         & $-10.83$ (0.46) \\
  06D2bk & 0.4990 (0.0010) & $ 0.94$ (0.28) & 1.05 (0.04) & $ 0.036$ (0.036) & 23.273 (0.039) &  8.617 (0.086) & $-14.88$ (0.21) & \nodata         & $-15.01$ (0.68) & \nodata         \\
  06D2cc & 0.5320 (0.0010) & $ 3.73$ (0.33) & 0.94 (0.04) & $ 0.089$ (0.046) & 23.468 (0.050) & 10.870 (0.108) & $-14.26$ (0.23) & \nodata         & $-14.73$ (0.68) & \nodata         \\
  06D2fb & 0.1242 (0.0004) & $ 1.80$ (0.14) & 0.96 (0.02) & $-0.004$ (0.020) & 19.772 (0.023) &  9.740 (0.124) & $-15.85$ (0.40) & \nodata         & $-16.22$ (0.70) & \nodata         \\
  06D3fp & 0.2680 (0.0010) & $-0.11$ (0.11) & 1.00 (0.02) & $ 0.104$ (0.019) & 21.748 (0.026) &  9.550 (0.259) & \nodata         & $-11.12$ (0.27) & \nodata         & $-11.12$ (0.46) \\
  06D3gh & 0.7200 (0.0050) & $-3.22$ (0.22) & 1.03 (0.04) & $-0.021$ (0.052) & 23.926 (0.056) &  9.300 (0.244) & $-15.34$ (0.05) & \nodata         & $-14.91$ (1.60) & \nodata         \\
  06D4cq & 0.4110 (0.0010) & $-0.25$ (0.10) & 1.04 (0.01) & $-0.005$ (0.020) & 22.562 (0.031) &  9.107 (0.088) & $-16.80$ (0.15) & \nodata         & $-16.74$ (0.66) & \nodata        

\enddata

\tablecomments{Uncertainties shown in parentheses.}
\tablenotetext{a}{Spectrum not used to determine adopted
maximum-brightness ejecta velocity.}

\end{deluxetable*}
\clearpage


\end{document}